\documentclass[11pt,a4paper]{article}

\usepackage[utf8]{inputenc}
\usepackage[T1]{fontenc}
\usepackage{lmodern}
\usepackage{textcomp}

\usepackage[margin=1in]{geometry}

\usepackage{graphicx}
\usepackage{booktabs}
\usepackage{multirow}
\usepackage{array}

\usepackage{amsmath,amssymb}
\usepackage{siunitx}

\usepackage[font=small,labelfont=bf]{caption}
\usepackage{subcaption}

\usepackage[numbers,sort&compress]{natbib}

\usepackage[hidelinks]{hyperref}

\usepackage[capitalise,nameinlink]{cleveref}

\crefname{figure}{Figure}{Figures}
\Crefname{figure}{Figure}{Figures}

\crefname{table}{Table}{Tables}
\Crefname{table}{Table}{Tables}

\crefname{section}{Section}{Sections}
\Crefname{section}{Section}{Sections}

\usepackage{float}

\usepackage{authblk}

\title{Metric Surface Reconstruction of Neurosurgical Scenes from Monocular Operating Microscope Images and Microscope Pose}

\author[1,2]{Thomas Bucher\thanks{Corresponding author: \href{mailto:thomas.bucher@unibe.ch}{thomas.bucher@unibe.ch}}}
\author[1,2]{Didier Neuenschwander}
\author[1,2]{Thomas Petutschnigg}
\author[2]{Michael Murek}
\author[2]{David Bervini}
\author[2]{Andreas Raabe}
\author[1,2]{Manuela Eugster}

\affil[1]{Neuro Robotics Group, ARTORG Center, University of Bern, Bern, Switzerland}
\affil[2]{Department of Neurosurgery, Inselspital, Bern University Hospital, University of Bern, Bern, Switzerland}

\date{}

\begin{document}

\maketitle

\begin{abstract}
\noindent\textbf{Objective:} The operating microscope records visual data during surgery, but its standard monocular video output is used only for documentation and contains no direct metric information. We evaluated whether a foundation-model-based reconstruction pipeline can recover the metric three-dimensional geometry of neurosurgical operative exposure from monocular microscope images, validated against laboratory-grade reference models.\\

\noindent\textbf{Methods:} A phantom-based laboratory study was performed using two aneurysm training phantoms in a navigated neurosurgical setup. Microscope images were acquired through the standard composite output of a ZEISS Pentero 800 microscope and stored synchronously with microscope pose information from Brainlab Cranial Navigation. Intrinsic and extrinsic calibration enabled operative surface reconstruction in the navigation coordinate system. Depth estimation was performed using the pretrained Depth Anything 3 model without task-specific fine-tuning. The resulting fused 3D point clouds were converted into continuous meshes using Poisson surface reconstruction. Reference surfaces were generated using structured-light surface scanning and fine-slice CT of the two phantoms used. Reconstruction accuracy, completeness, and F-score were evaluated against the corresponding reference geometries.\\

\noindent\textbf{Results:} For phantom A, representing a deeper retractor-defined corridor, reconstruction accuracy ranged from 1.95 $\pm$ 1.70 mm to 2.33 $\pm$ 2.15 mm across the evaluated image subsets. For phantom B, representing a more directly exposed surface, reconstruction accuracy ranged from 1.02 $\pm$ 0.93 mm to 1.52 $\pm$ 1.21 mm. Reconstructions using larger image sets were associated with improved reconstruction completeness, whereas reconstruction accuracy remained within a narrower range. Qualitative corridor analysis demonstrated preservation of overall corridor geometry with local deviations in regions of incomplete reconstruction.\\

\noindent\textbf{Conclusions:} Standard monocular microscope images combined with navigation-derived pose data allow to reconstruct millimeter-range three-dimensional surfaces using a foundation-model-based pipeline. The findings establish technical feasibility in a controlled phantom setting and provide a basis for translating the approach toward objective quantification of operative exposures, fusion with preoperative imaging, and the dimensional characterization of working spaces for future surgical instrumentation.
\end{abstract}


\newpage

\section{Introduction}

The operating microscope has shaped cranial neurosurgery since its first cranial application in 1957~\cite{uluc_operating_2009}. While operating microscopes provide stereoscopic vision to the surgeon through binocular eyepieces, the standard image output, used for recording and documentation, is monocular. Throughout every surgical procedure, the operating microscope captures rich visual data. However, this data remains qualitative. Surgeons intuitively interpret geometrical information such as distances, surface geometry, and surgical corridor dimensions, yet these metrics are not measured. As a result, valuable quantitative information about the surgical environment and procedure remains inaccessible. Photogrammetric methods, which reconstruct three-dimensional surfaces from sets of overlapping two-dimensional images, have been applied to address this gap, predominantly in cadaver studies and for educational purposes where visual fidelity has been prioritized over metric accuracy~\cite{de_benedictis_photogrammetry_2018, hanalioglu_development_2022, krogager_intraoperative_2024}. None of these methods have been quantitatively validated against precise ground-truth measurements. The ability to accurately reconstruct geometrical information of operative surfaces and corridors from data already captured during surgery would enable extracting objective metrics for surgical assessment and research. This capability is expected to become increasingly relevant in the era of minimally invasive approaches~\cite{lan_international_2021}.

A potential technical pathway for the extraction of objective intraoperative geometry information has emerged through recent advances in deep learning, particularly with the availability of open-source foundation models. Unlike conventional deep learning models, which are trained for specific tasks using dedicated datasets, foundation models are trained on large and diverse image collections, enabling their application to new domains with limited or even no task-specific adaptation.

Monocular images acquired from the surgical microscope do not contain sufficient geometric information to enable depth estimation using classical computer vision methods. However, when combined with neuronavigation data describing microscope poses, foundation models could overcome this limitation by incorporating learned visual priors, i.e. implicit knowledge about scene layout, object shape, and texture, to estimate depth from monocular views. This kind of depth recovery from monocular surgical imaging has previously been demonstrated in endoscopic imaging~\cite{liu_dense_2020, zanier_real-time_2026, tong_real--virtual_2021}. However, these studies have focused on tubular environments such as sinus and transsphenoidal endoscopy and have not yet been translated to the more complex setting of open microsurgical neurosurgery.

The goal of this study was to investigate if monocular images from the operating microscope combined with neuronavigation data are sufficient to reliably reconstruct the geometry of operative surfaces and surgical corridors using a foundation-model-based reconstruction pipeline. As a controlled pilot, we applied our method in a laboratory setting using commercially available hyper-realistic pterional approach aneurysm surgery phantoms. We obtained reference geometries (i.e., reference models) through high-precision optical surface scanning and fine-slice computed tomography (CT) of the phantoms, which served as a comparison for our foundation-model-based reconstruction pipeline.

\newpage
\section{Methods}

We performed a phantom-based laboratory study to demonstrate the technical feasibility of the proposed approach. Therefore, no institutional or ethical review or approval was required. No formal clinical reporting guideline was applicable to this technical feasibility study.

\subsection{Experimental setup and data acquisition}

\begin{figure}[!htbp]
    \centering
    \includegraphics[width=\linewidth]{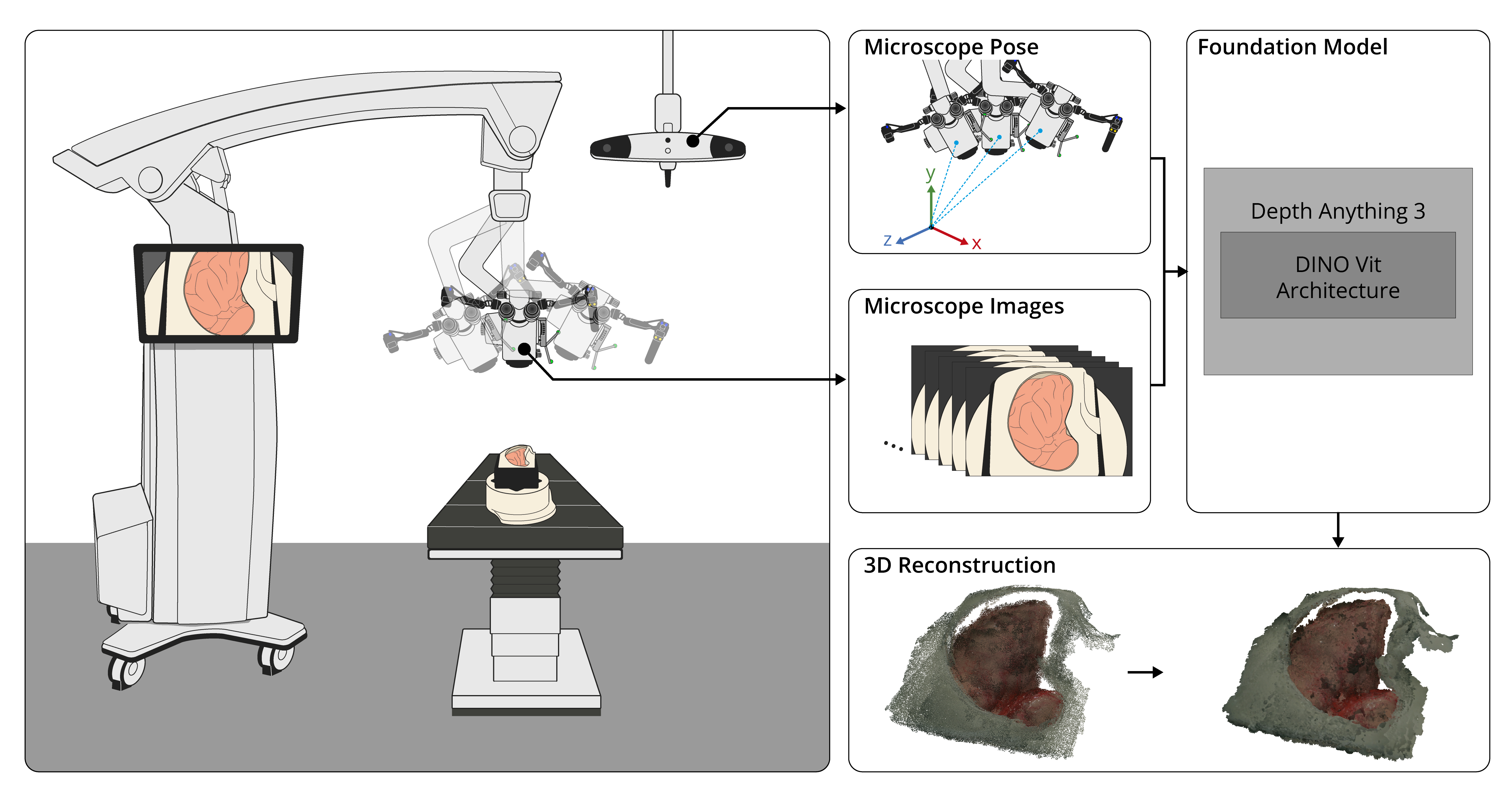}
    \caption{Experimental setup and vision pipeline. Microscope images and corresponding microscope pose information were acquired synchronously from aneurysm training phantoms using a navigated surgical microscope. These data were provided as input to Depth Anything 3 (DA3), which uses a DINO vision transformer backbone and generated dense geometric output in the form of a point cloud. The point cloud was subsequently converted into a continuous surface reconstruction for quantitative analysis.}
    \label{fig:experimental_setup}
\end{figure}

All experiments were conducted in a controlled laboratory setting, replicating a navigated neurosurgical workflow. A surgical microscope (Pentero 800, Carl Zeiss Meditec, Oberkochen, Germany) was integrated with a neuronavigation system (Brainlab Cranial Navigation, version 3.1.5.177; Brainlab, Munich, Germany). Both systems were connected to a dedicated recording workstation, enabling synchronous acquisition of microscope images and corresponding microscope pose information. An overview of the experimental setup and vision-based reconstruction workflow is shown in \cref{fig:experimental_setup}.

The microscope image stream was acquired through the dedicated composite output of the surgical microscope at a resolution of 720 $\times$ 576 pixels in the PAL UYVY format. Navigation data were obtained through IGTLink, a research interface provided by Brainlab. For each image acquisition, microscope images were stored synchronously with the corresponding microscope pose in the navigation coordinate system.

Two commercially available aneurysm training phantoms were used to validate technical feasibility of the approach: the AneurysmBox phantom (UpSurgeOn, Milan, Italy; hereafter, phantom A) and the NeuraBubble phantom (SurgeonsLab, Bern, Switzerland; hereafter, phantom B). For phantom A, a neurosurgeon placed a retractor to expose the depicted medial cerebral artery aneurysm in a manner representative of a standard surgical approach.

We performed two experiments, one for each phantom, using a fixed microscope zoom/focus setting. For each experiment microscope images and corresponding microscope pose information were acquired for multiple microscope poses around the target region, i.e. the phantom surface. Because of differences in phantom size, 50 images were acquired for phantom A and 36 images for phantom B. The number of images per phantom was defined during preliminary testing based on the number of views required to obtain sufficient visual coverage of the entire phantom surface. To evaluate the influence of the number of input views on the surface reconstruction quality, additional reconstructions were generated from subsets of the acquired images. For phantom A, subsets of 5, 10, 20, 30, and 40 images were evaluated; for phantom B, subsets of  5, 10, 20, and 30 images were evaluated. Images for each subset were selected at approximately uniform intervals across the acquisition sequence to preserve coverage of the target region. Because only discrete image subsets were evaluated, this analysis was intended to assess observed performance trends rather than establish a continuous or linear relationship between image number and reconstruction quality.

Images were acquired at microscope zoom settings of 3.5. The corresponding focus setting was determined by the zoom setting and working distance and kept constant during each acquisition. Microscope illumination was kept constant at 5\% for all recordings.

\subsection{Calibration method}

Before image acquisition, the microscope and neuronavigation system were calibrated according to the clinical workflow recommended by Brainlab. This included registration of the tracked microscope within the Brainlab navigation system.

For the surgical microscope camera, an intrinsic camera calibration was performed using Zhang’s method based on multiple microscope images of a planar checkerboard calibration target~\cite{zhang_flexible_2000}. For the calibration acquisitions, a checkerboard pattern with 9 $\times$ 6 squares and a square size of 2 mm was used. Focal length and principal point were estimated by minimizing the reprojection error between detected checkerboard corners and their known planar coordinates.

Extrinsic microscope camera calibration was performed to determine the rigid transformation from the microscope camera coordinate system to the navigation coordinate system. For this purpose, corresponding observations of the calibration target were acquired together with the tracked microscope poses. The hand–eye transformation was estimated by solving for the rigid transformation that best aligned the camera-derived calibration target pose with the corresponding pose in the navigation coordinate system. This transformation enabled the back-projected depth maps and reconstructed point clouds to be expressed in a common metric coordinate frame.

\subsection{Phantom reference surface}

The reference surface for validation of the surface reconstruction method was acquired using two complementary modalities, one surface-based and one volumetric. For Phantom A, both reference acquisitions were performed with the retractor in place.

Directly accessible surface geometries were captured using a structured-light 3D scanner (GOM ATOS Core 100, Carl Zeiss GOM Metrology GmbH, Braunschweig, Germany), which provides point spacings of approximately 0.02–0.08 mm depending on the measuring volume.

CT imaging was performed as a complementary volumetric reference using a SOMATOM X.ceed scanner (Siemens Healthineers, Erlangen, Germany). Acquisition parameters included 120 kVp, 45 mA, 82 mAs, 0.5 mm slice thickness, 0.448 $\times$ 0.448 mm in-plane pixel spacing, and a 512 $\times$ 512 matrix. Images were reconstructed using an Hr40s kernel, resulting in a voxel size of 0.448 $\times$ 0.448 $\times$ 0.5 mm.

Because both modalities represent approximations of the true phantom geometry rather than absolute physical ground truth, structured-light and CT-derived reference surfaces were compared to characterize their agreement and identify regions in which modality-specific limitations could influence subsequent reconstruction evaluation. The structured-light surface scan and CT-derived segmentation model were first registered using iterative closest point (ICP) registration~\cite{besl_method_1992}. A directed comparison was then performed from the structured-light surface scan to the CT-derived segmentation model. The structured-light surface scan was uniformly sampled, and each sampled point was compared with the CT-derived surface using the closest point-to-surface distance. Deviations were reported as mean $\pm$ standard deviation (SD) surface distance.

For quantitative reconstruction evaluation, the acquired phantom reference surfaces were registered to the reconstructed surfaces using ICP registration. The CT-derived surface was used as the reference surface for phantom A, whereas the structured-light surface scan was used as the reference surface for phantom B. This selection reflected the different visual accessibility of the relevant target regions: phantom A included deeper corridor structures, whereas phantom B provided directly accessible surface geometry.

\subsection{Reconstruction pipeline}

Dense surface reconstruction, i.e. depth maps, were estimated from the microscope images and corresponding microscope poses using Depth Anything 3 (DA3)~\cite{lin_depth_2025}, a deep learning–based foundation model for monocular and multi-view visual geometry estimation. DA3 uses a DINOv2-based vision transformer backbone~\cite{oquab_dinov2_2024}, a self-supervised visual feature representation designed to learn robust image features without manual annotation. DA3 can incorporate known camera poses and directly combines information from multiple views to generate metric depth estimates. In this study, the recorded microscope poses were provided to DA3 together with the corresponding microscope images, allowing the predicted surface geometry to be reconstructed in a metric coordinate system. The per-view geometric predictions were therefore not treated as independent reconstructions; instead, DA3 generated a multi-view-consistent output, which was exported as a fused metric 3D point cloud of the observed target. A continuous mesh was generated from the fused 3D point cloud using Poisson surface reconstruction~\cite{kazhdan_poisson_2006}, a method that estimates a smooth surface from oriented 3D points.

A publicly available DA3 model (DA3NESTED-GIANT-LARGE with 1.4 billion parameters) was used with pretrained weights. The DA3 models were trained exclusively on public datasets, including general computer-vision datasets of synthetic scenes, indoor environments, and multi-view reconstructions~\cite{lin_depth_2025}. Therefore, the model was not trained on any medical datasets and no fine-tuning of the model parameters to our experimental setup was performed. All computations were performed offline after image acquisition using a custom Python-based reconstruction pipeline.

\subsection{Surface reconstruction evaluation metrics}

Reconstruction quality was evaluated by comparing the reconstructed surfaces with the corresponding reference surface. To avoid bias from differences in mesh resolution or triangle density, all surface comparisons were performed using area-weighted uniform sampling of the triangular meshes. Points were sampled with probability proportional to triangle area, resulting in an approximately uniform distribution over the surface. For reconstruction accuracy assessment, sampled points from the reconstructed surface were compared with the reference surface using the closest point-to-surface distance. These distances were reported as reconstruction accuracy using mean $\pm$ SD and 95th percentile error. Reconstruction completeness was assessed by applying the inverse comparison, in which sampled points from the reference surface were compared with the reconstructed surface using the closest point-to-surface distance. This quantified how completely the reconstructed surface covered the reference surface. Because accuracy and completeness were computed as directed surface distances in opposite directions, the two metrics capture different reconstruction properties and are not expected to be symmetric. A reconstruction may therefore show low accuracy error for the surface regions that were reconstructed, while still showing higher completeness error if parts of the reference surface were not recovered.

Color-coded surface deviation maps were generated to visualize the spatial distribution of reconstruction accuracy. For this purpose, distances from the reconstructed surface to the corresponding reference surface were computed and mapped onto the reconstructed surface using a color scale in millimeters.

Threshold-based reconstruction quality was evaluated using the F-score~\cite{knapitsch_tanks_2017}, a standard comparison metric for surface reconstruction, which summarizes both local reconstruction accuracy and surface coverage in a single value. F-scores were computed using a 1-mm distance threshold ($F_1$).

A task-specific corridor analysis was performed to assess whether the reconstructed geometry could support direct metric measurements within a surgical cavity or corridor. Representative point-to-point distances were therefore measured on the reconstructed and reference surfaces using the Measuring Tool in MeshLab (version 2020.09; Visual Computing Lab, ISTI-CNR, Pisa, Italy).

\newpage
\section{Results}

\subsection{Phantom reference surface comparison}

\begin{figure}[!htbp]
    \centering
    \includegraphics[width=\linewidth]{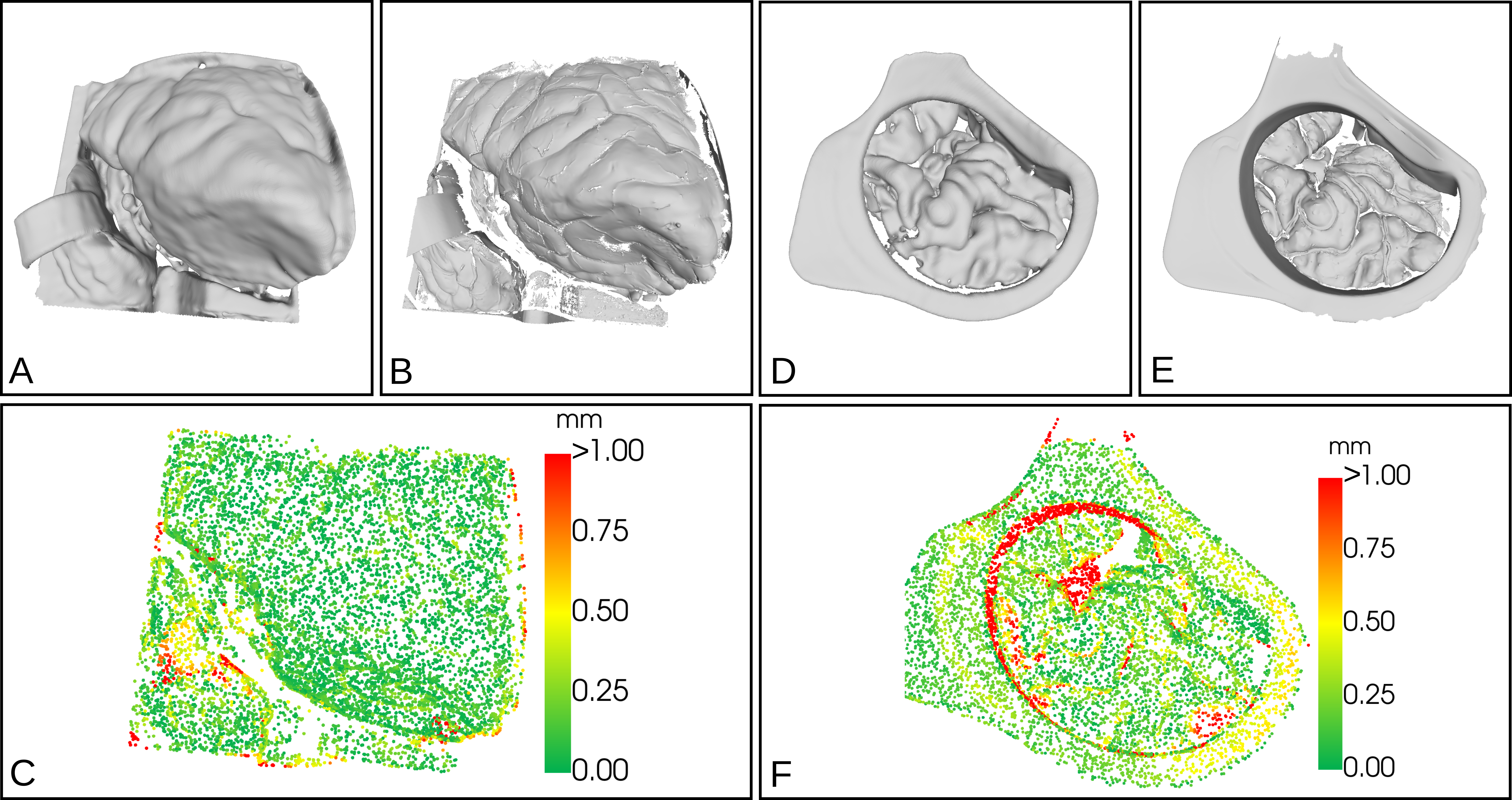}
    \caption{Comparison of phantom reference models. A–C: Phantom A. D–F: Phantom B. A and D: CT-derived segmentation models. B and E: Structured-light surface scan-based models. C and F: Color-coded surface deviation maps showing the difference between the structured-light surface scan and the CT-derived segmentation model after registration. The color scale was capped at 1 mm; deviations greater than 1 mm are shown with the maximum color value.}
    \label{fig:phantom_reference_models}
\end{figure}

Representative 3D reference surfaces generated from CT segmentation and surface scanning are shown for both phantoms in \cref{fig:phantom_reference_models}. In the directed comparison from the surface scan to the CT-derived segmentation model, the mean surface deviation was 0.188 $\pm$ 0.298 mm for phantom A and 0.487 $\pm$ 0.758 mm for phantom B. The largest deviations were observed in deep cavity regions and narrow recesses, consistent with incomplete surface coverage by line-of-sight optical scanning.

\subsection{Qualitative reconstruction pipeline output}

\begin{figure}[!htbp]
    \centering
    \includegraphics[width=0.8\linewidth]{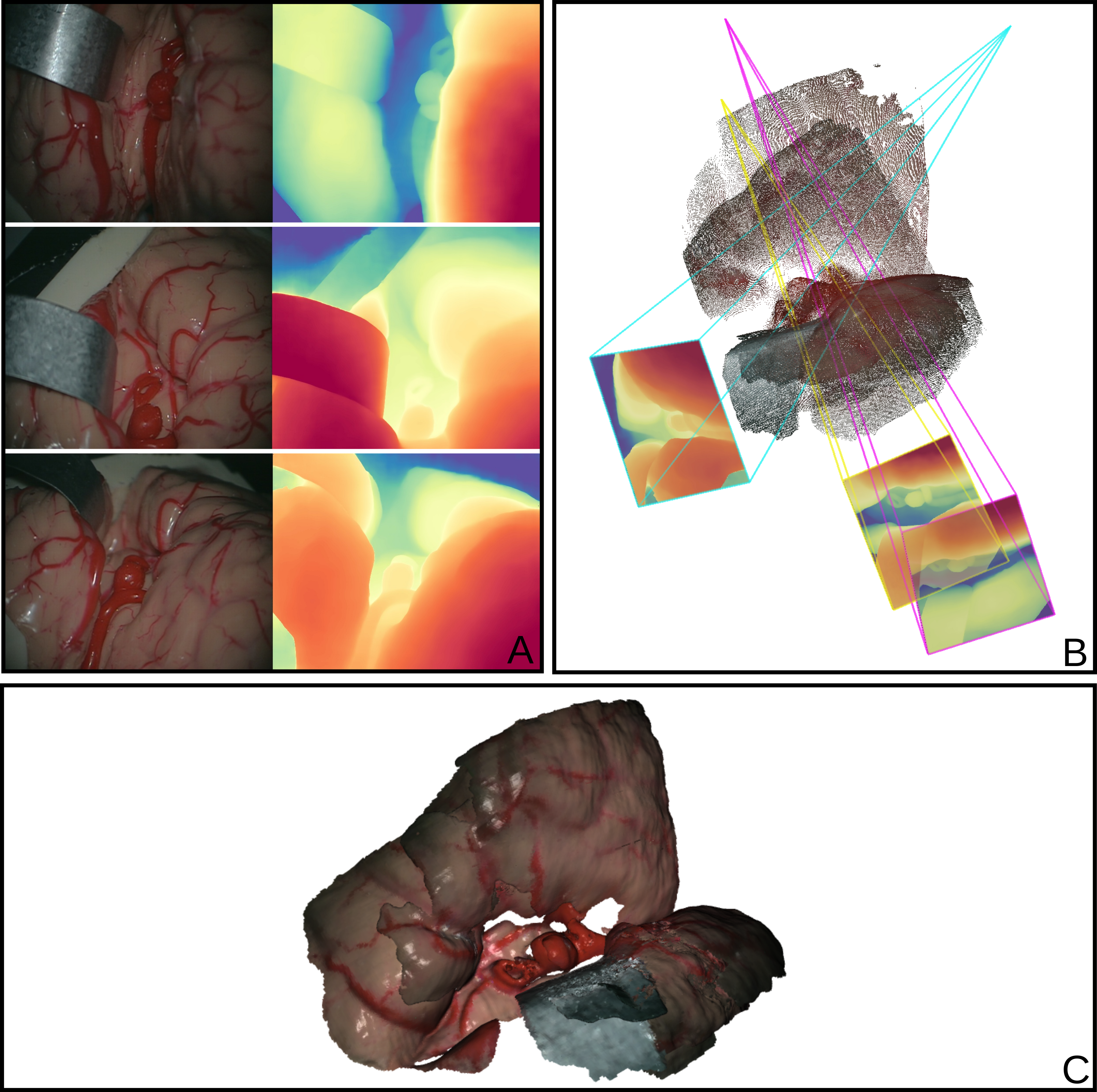}
    \caption{Qualitative reconstruction pipeline output for phantom A using a subset of three input images. A: Representative microscope images and corresponding DA3-estimated dense depth maps. B: Recorded microscope poses and reconstructed point cloud generated from the DA3 output after transformation into the navigation coordinate system using calibration and pose information. C: Continuous surface mesh generated from the reconstructed point cloud using Poisson surface reconstruction.}
    \label{fig:qualitative_pipeline_output}
\end{figure}

Representative outputs of the reconstruction pipeline are shown in \cref{fig:qualitative_pipeline_output}. The generated depth maps reproduced the overall spatial configuration of the visible phantom surfaces and corridor geometries. The fused point clouds provided dense geometric representations of the visible scene, and Poisson surface reconstruction generated continuous meshes for subsequent quantitative analysis.

Local artifacts, such as small mesh irregularities, duplicated surface layers, and incomplete surface patches, were observed in regions with low illumination and in areas with substantial overlap across multiple input images. These artifacts were most commonly associated with local inconsistencies in the fused point cloud and incomplete surface coverage rather than complete reconstruction failure.

\subsection{Quantitative surface reconstruction accuracy}

\begin{figure}[p]
    \centering

    \includegraphics[width=0.8\linewidth]{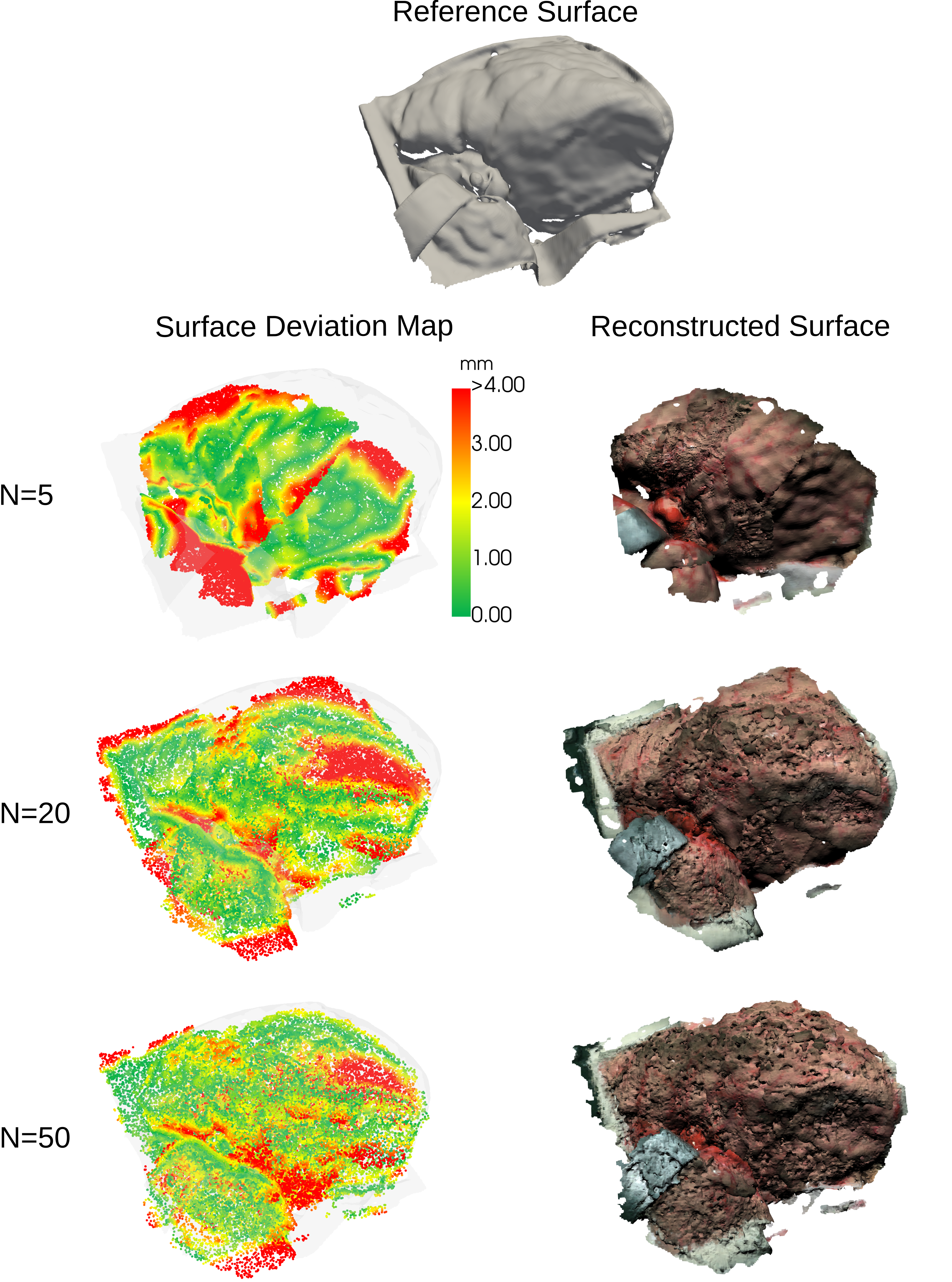}
    \caption{Surface reconstruction accuracy for phantom A. Qualitative comparison of the reconstructed surface with the reference surface for phantom A at a fixed microscope zoom setting of 3.5 using subsets of N = 5, 20, and 50 input images. Surface deviation maps are shown with the color scale capped at 4 mm; deviations greater than 4 mm are shown with the maximum color value.}
    \label{fig:accuracy_phantom_a}

    \vspace{1.5em}

    \captionof{table}{Quantitative surface reconstruction performance for phantom A at a fixed microscope zoom setting of 3.5 using different subsets of input images.}
    \label{tab:phantom_a}

    \small
    \begin{tabular}{rrrrrr}
    \toprule
    Number of Images & \multicolumn{2}{c}{Accuracy (mm)} & \multicolumn{2}{c}{Completeness (mm)} & $F_1$ (\%) \\
    \cmidrule(lr){2-3}\cmidrule(lr){4-5}
     & Mean$\pm$SD & P95 & Mean$\pm$SD & P95 & \\
    \midrule
    5 & 2.03$\pm$1.99 & 6.47 & 7.57$\pm$9.11 & 27.71 & 27.3 \\
    10 & 2.33$\pm$2.15 & 6.51 & 3.67$\pm$3.59 & 11.19 & 32.4 \\
    20 & 2.03$\pm$1.69 & 5.30 & 2.78$\pm$3.23 & 9.65 & 36.3 \\
    30 & 2.12$\pm$1.72 & 5.57 & 2.56$\pm$2.96 & 8.90 & 36.2 \\
    40 & 2.07$\pm$1.66 & 5.39 & 1.97$\pm$2.78 & 7.30 & 39.9 \\
    50 & 1.96$\pm$1.55 & 5.06 & 2.05$\pm$3.07 & 7.79 & 40.8 \\
    \bottomrule
    \end{tabular}

\end{figure}

\begin{figure}[p]
    \centering

    \includegraphics[width=0.8\linewidth]{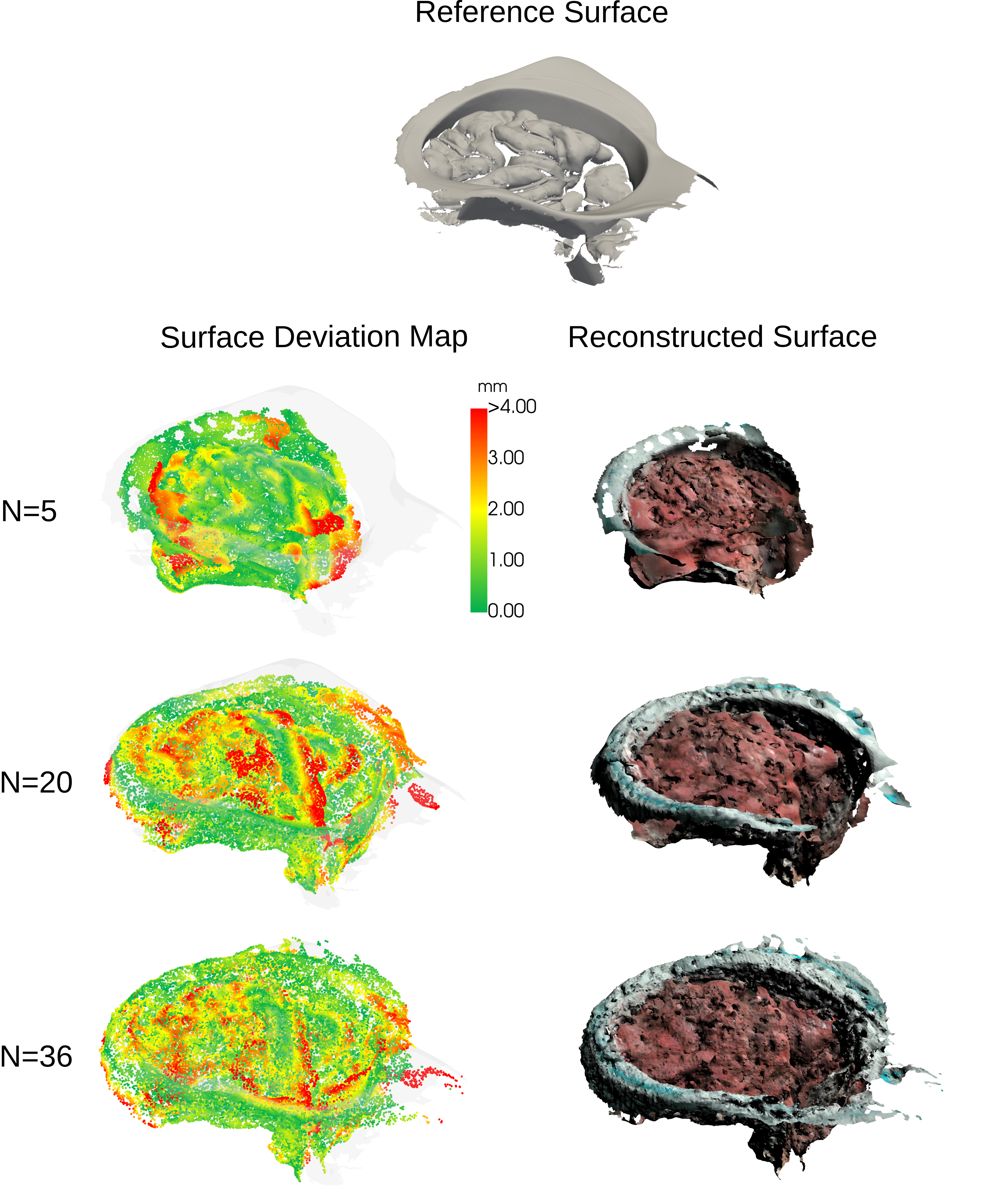}
    \caption{Surface reconstruction accuracy for phantom B. Qualitative comparison of the reconstructed surface with the reference surface for phantom B at a fixed microscope zoom setting of 3.5 using subsets of N = 5, 20, and 36 input images. Surface deviation maps are shown with the color scale capped at 4 mm; deviations greater than 4 mm are shown with the maximum color value.}
    \label{fig:accuracy_phantom_b}

    \vspace{1.5em}

    \captionof{table}{Quantitative surface reconstruction performance for phantom B at a fixed microscope zoom setting of 3.5 using different numbers of input images.}
    \label{tab:phantom_b}

    \small
    \begin{tabular}{rrrrrr}
    \toprule
    Number of Images & \multicolumn{2}{c}{Accuracy (mm)} & \multicolumn{2}{c}{Completeness (mm)} & $F_1$ (\%) \\
    \cmidrule(lr){2-3}\cmidrule(lr){4-5}
     & Mean$\pm$SD & P95 & Mean$\pm$SD & P95 & \\
    \midrule
    5 & 1.02$\pm$0.93 & 2.92 & 3.34$\pm$4.63 & 13.88 & 56.4 \\
    10 & 1.52$\pm$1.21 & 3.86 & 2.29$\pm$3.43 & 11.00 & 48.9 \\
    20 & 1.34$\pm$1.11 & 3.35 & 1.01$\pm$1.58 & 4.27 & 58.3 \\
    30 & 1.51$\pm$1.23 & 3.90 & 0.87$\pm$1.25 & 3.33 & 56.1 \\
    36 & 1.39$\pm$1.15 & 3.33 & 0.65$\pm$0.72 & 1.81 & 57.9 \\
    \bottomrule
    \end{tabular}

\end{figure}

Surface deviation maps for the reconstructions acquired at a fixed microscope zoom setting of 3.5 are shown for phantom A in \cref{fig:accuracy_phantom_a} and for phantom B in \cref{fig:accuracy_phantom_b}. Quantitative reconstruction metrics are summarized in \cref{tab:phantom_a} for phantom A and \cref{tab:phantom_b} for phantom B.

For phantom A, the reconstruction accuracy ranged from 1.95 $\pm$ 1.70 mm (mean $\pm$ SD) to 2.33 $\pm$ 2.15 mm across the evaluated image subsets. The 95th percentile accuracy error ranged from 5.06 mm to 6.51 mm. Across the predefined image subsets, completeness was generally better for larger subsets, with mean completeness ranging from 7.57 $\pm$ 9.11 mm using 5 images to 1.97 $\pm$ 2.78 mm using 40 images. The F-score ranged from 27.3\% to 40.8\%, with the highest F-score ($F_1$) observed using 50 input images.

For phantom B, the reconstruction accuracy ranged from 1.02 $\pm$ 0.93 mm (mean $\pm$ SD) to 1.52 $\pm$ 1.21 mm. The 95th percentile accuracy error ranged from 2.92 mm to 3.90 mm. Across the predefined image subsets, completeness was generally better for larger subsets, with mean completeness ranging from 3.34 $\pm$ 4.63 mm using 5 images to 0.65 $\pm$ 0.72 mm using 36 images. The F-score ranged from 48.9\% to 58.3\%, with the highest F-score observed using 20 input images.

Across both phantoms, the difference between accuracy and completeness indicates that reconstructed regions were locally close to the reference surface, whereas unreconstructed or incompletely covered regions contributed mainly to the completeness error.

\newpage
\subsection{Corridor subregion analysis}

To illustrate reconstruction performance in a clinically relevant anatomical region, a corridor subregion of phantom A was visualized separately. The region included the retractor-exposed surgical corridor toward the aneurysm target and was reconstructed using the previously evaluated 5-image subset; the corresponding quantitative metrics are reported in \cref{tab:phantom_a}.

\begin{figure}[!htbp]
    \centering
    \includegraphics[width=0.6\linewidth]{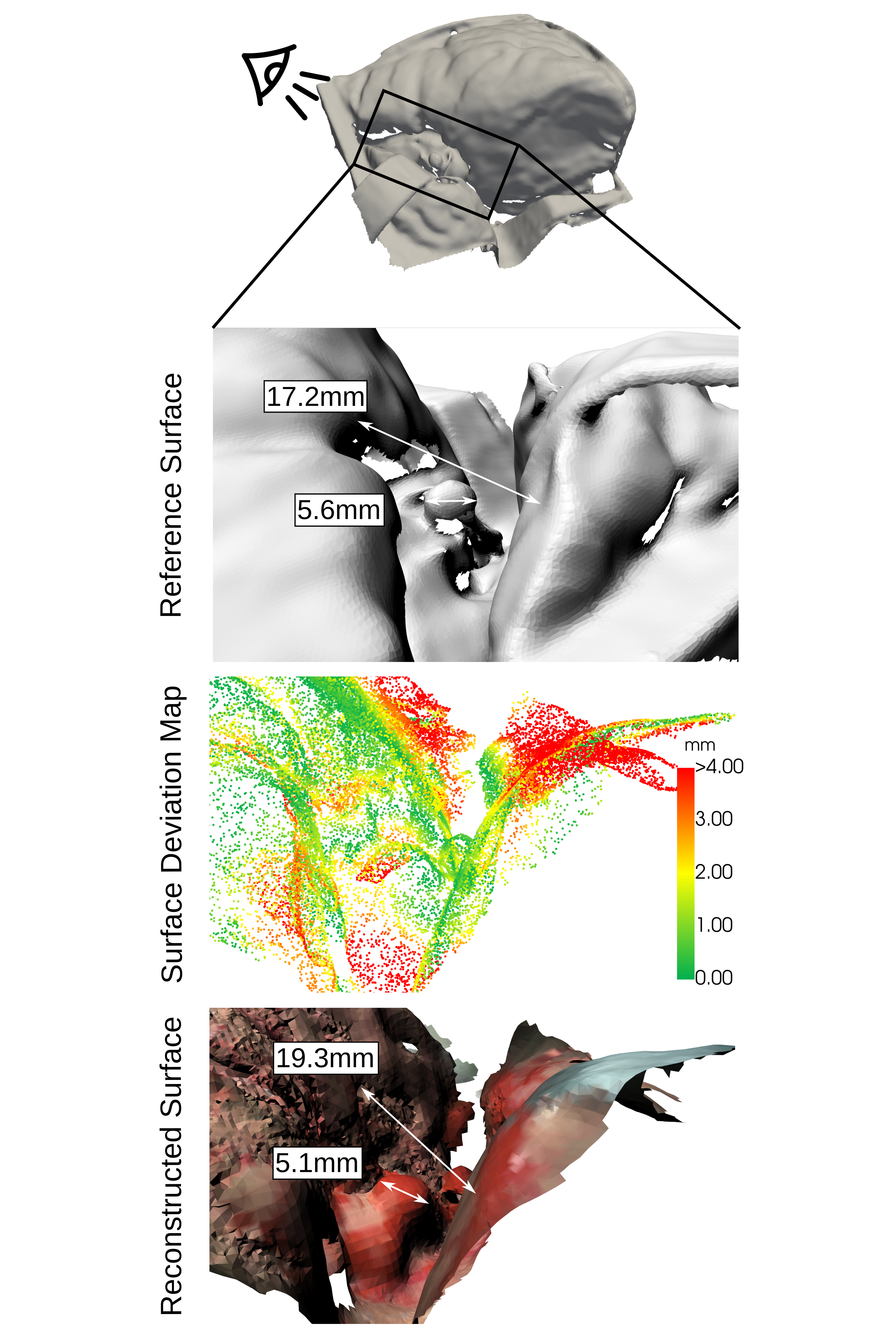}
    \caption{Surface reconstruction accuracy for the corridor subregion of phantom A. Comparison of the reconstructed surface with the reference surface for phantom A at a fixed microscope zoom setting of 3.5 using a subset of N = 5 input images. Surface deviation maps are shown with the color scale capped at 4 mm; deviations greater than 4 mm are shown with the maximum color value.}
    \label{fig:corridor_subregion}
\end{figure}

\cref{fig:corridor_subregion} shows the reference surface, reconstructed surface, surface deviation map, and two representative distance measurements performed on both the reference and reconstructed surfaces. The visualization demonstrates preservation of the overall corridor geometry, with local deviations in regions of incomplete reconstruction and limited geometric coverage.

\newpage
\section{Discussion}

This study shows that a foundation-model-based pipeline can reconstruct three-dimensional surfaces and operative corridors from standard monocular microscope images, with accuracy in the millimeter range. In phantom B, which represented an exposed anatomical surface, the mean reconstruction accuracy reached 1.02 mm, and up to 58.3\% of the reconstructed surface laid within 1 mm of the reference (F-score). In phantom A, which represented a deeper corridor with retraction-defined boundaries, accuracy ranged from 1.95 to 2.33 mm and F-scores from 27.3\% to 40.8\%. Consistent with our expectation, open surfaces close to the microscope were reconstructed more accurately than deeper corridor geometries.

\subsection{Results of the applied reconstruction pipeline}

The reconstruction pipeline generated metric 3D point clouds and surface meshes from monocular microscope images combined with corresponding microscope poses. Across both phantoms, larger image subsets were primarily associated with improved completeness, whereas accuracy values remained within a narrower range, suggesting that additional views mainly increased surface coverage rather than uniformly improving local surface accuracy. Because only predefined discrete subsets were evaluated, these findings should be interpreted as observed trends rather than evidence of a continuous or linear relationship between the number of input images and reconstruction quality. Surface deviation maps showed spatially heterogeneous errors, with larger deviations observed in regions of incomplete local reconstruction and limited geometric coverage.

The exposed surface of phantom B was reconstructed more accurately than the deeper corridor of phantom A, where lower illumination, oblique viewing angles, and incomplete local coverage created more challenging conditions. The F-scores at the 1-mm threshold, reaching up to 58.3\% for phantom B and 40.8\% for phantom A, indicate that a substantial portion of the reconstructed surface deviates from the reference surface by more than 1 mm. Since we applied a generic foundation model with pretrained weights to a surgical imaging task it was not explicitly designed for, the achieved millimeter-scale reconstruction accuracy is encouraging and indicates substantial room for improvement through targeted adaptation and fine-tuning.

\subsection{Comparison with existing approaches}

Liu et al. estimated depth from monocular sinus endoscopy and reported submillimeter residual error against patient CT scans, validated through point correspondences in two patients~\cite{liu_dense_2020}. Tong et al. achieved a mean absolute error of about 1.1 mm at single target points in a transnasal phantom~\cite{tong_real--virtual_2021}, and Zanier et al. showed that real-time depth estimation is feasible during transsphenoidal pituitary surgery~\cite{zanier_real-time_2026}.  These studies were conducted in tubular environments characterized by relatively uniform geometry, controlled illumination, and validation against sparse reference measurements. In addition, depth estimation was performed on a frame-by-frame basis, without integrating the results into a unified quantitative three-dimensional surface reconstruction. In contrast, our approach achieved errors within the same order of magnitude while reconstructing a complete surface in an open surgical geometry, representing a substantially more challenging setting.

Within cranial neurosurgery, photogrammetric methods have been proposed to reconstruct 3D surfaces from overlapping 2D images using classical feature detection and triangulation~\cite{de_benedictis_photogrammetry_2018, hanalioglu_development_2022, krogager_intraoperative_2024}. For example, a recent study by Sinosi and Rubio compared photogrammetry in cadaveric pterional corridors against neuronavigation, reporting sub-millimeter mean bias~\cite{sinosi_open_2025}. However, performance of photogrammetric methods depends on rich texture, high image overlap, and consistent illumination, conditions that are difficult to achieve in the operating room. Furthermore, the methods require dedicated acquisitions that interrupt the surgical workflow. To our knowledge, a surgical surface reconstruction method that can be integrated disturbance-free into the microneurosurgical workflow has not previously been reported.

\subsection{Future applications}

Neurosurgery has consistently expanded the capabilities of the microscope through complementary technologies, ranging from intraoperative fluorescence imaging to navigated visualization~\cite{raabe_near-infrared_2003, stummer_fluorescence-guided_2006, bopp_augmented_2022}. Rather than introducing additional hardware, the present approach enables geometric interpretation of the existing image stream. What evolves is not the microscope but the information that can be extracted from it:

\subsubsection{Anatomical surfaces}

In the surface-focused setting represented by phantom B, the reconstructed surface anatomy reached an accuracy of 1.02 mm against the structured-light scanned reference surface, approaching the order of magnitude relevant for clinical surface registration. If translated to the operating room, this level of fidelity could provide a natural registration target for fusion with preoperative imaging modalities such as MRI and CT. In this way, the approach could contribute to brain-shift compensation through optico-neuronavigation updates.

\subsubsection{Surgical corridor}

The surgical corridor itself currently lacks an objective measure suitable for surgical studies. Craniotomy size and skin incision length are commonly reported as proxies for minimally invasive surgery~\cite{lan_international_2021, jagersberg_quantification_2017}, but these describe only the outer boundary of the approach and not the parenchymal pathway that determines exposure and invasiveness. The dissected corridor, the available instrument manipulation space, and the amount of brain retraction are better measures but remain almost entirely underreported. The reconstruction accuracy at depth observed for phantom A, in the 2 mm range, is coarser than the surface accuracy of phantom B but reflects the more challenging conditions of deeper corridors. Even at this accuracy, the reconstructed corridor reproduces boundary geometry from entry to depth in a way that existing approaches do not. The Volume of Operative Maneuverability~\cite{tariciotti_volume_2025}, for example, captures the corridor by probing a small set of points with the standard neuronavigation pointer and fitting them to ellipsoid primitives. The methodology is rigorous, but the corridor is reduced to a parametric approximation rather than captured directly, and acquisition requires dedicated intraoperative steps. With a direct corridor metric available, exposure and invasiveness could be compared objectively across surgeons, centers, and surgical philosophies, providing a reproducible basis for outcome reporting that has so far depended on qualitative description. The same measurements would also define the operative envelope within which next-generation instrumentation and surgical robotics will need to function.

\subsection{Future development directions}

First, domain-specific training of the foundation model, combining phantom-based and intraoperative microscope images, is expected to improve reconstruction quality. Second, continuous reconstruction over the duration of surgery would enable real-time intraoperative use, requiring model compression, dedicated hardware, and integration with the navigation system. Stereoscopic acquisition with 3D microscopes would in principle simplify reconstruction~\cite{kumar_persistent_2015}, but 3D-capable systems are reported to be available in only about one third of neurosurgical departments~\cite{khafaji_use_2025}. Third, processing of archived surgical recordings with corresponding microscope pose data would enable retrospective analysis of surgical corridor dimensions.

\subsection{Limitations}

This is a phantom-based pilot study with several limitations. The training phantoms do not represent the optical and mechanical properties of living tissue, and validation in cadaveric and in vivo settings will be required before clinical conclusions can be drawn. Image acquisition was performed under controlled conditions, with fixed zoom, dim illumination, and deliberate microscope motion to ensure sufficient multi-view coverage. Whether comparable reconstruction quality can be achieved from intraoperatively acquired images remains to be demonstrated. The reconstruction reflects a single point in time, while surgical exposure evolves continuously, a limitation shared with pointer-based and photogrammetric methods. The foundation model was used with pretrained weights without domain-specific fine-tuning, which supports reproducibility. Image acquisition relied on the standard PAL composite output (720 $\times$ 576 pixels) chosen for cross-vendor generalizability, a conservative imaging baseline that modern higher-resolution capture is likely to improve.

\section{Conclusions}

The pipeline based on an open-source foundation model can reconstruct three-dimensional surfaces and operative corridors from standard monocular microscope images with millimeter-range accuracy measured against phantom-based reference surfaces. The approach requires no dedicated stereoscopic hardware and remains compatible with any conventional surgical microscope. If translated to the operating room, it could provide a quantitative basis for objective comparison of surgical exposures, integration with preoperative imaging, and the dimensional requirements of future surgical instrumentation and robotics.

\section*{Acknowledgments}

We thank Prof. Dr. Jürgen Burger and Daniel Widmer from the School of Biomedical and Precision Engineering for their support in performing the optical scanning of the phantoms. We also thank Anja Giger for her assistance with the medical illustration.

\newpage
\bibliographystyle{IEEEtranN}
\bibliography{references}

\end{document}